# Quantum register based on structured diamond waveguide with NV centers

Alexander V. Tsukanov*, Igor Yu. Kateev, Alexander A. Orlikovsky

Institute of Physics and Technology, Russian Academy of Sciences, 117218, Nakhimovsky pr. 34, Moscow, Russia

## ABSTRACT

We propose a scheme of quantum information processing with NV-centers embedded inside diamond nanostructure. Single NV-center placed in the cavity plays role of an electron spin qubit which evolution is controlled by microwave pulses. Besides, it couples to the cavity field via optical photon exchange. In their turn, neighbor cavities are coupled to each other through the photon hopping to form a bus waveguide mode. This waveguide mode overlaps with all NV-centers. Entanglement between distant centers is organized by appropriate tuning of their optical frequency relative to the waveguide frequency via electrostatic control without lasers. We describe the controlled-Z operation that is by one order of magnitude faster than in off-resonant laser-assisted schemes proposed earlier. Spectral characteristics of the one-dimensional chain of microdisks are calculated by means of numerical modeling, using the approach analogous to the tight-binding approximation in the solid-state physics. The data obtained allow to optimize the geometry of the microdisk array for the effective implementation of quantum operations.

## 1. INTRODUCTION

Structure defects in diamond, the so-called color centers, attract much attention due to their unique magnetic and optical properties. To date, substitution nitrogen with neighbor vacancy in diamond lattice (NV-center) is the most studied of them[1-3]. Its negatively charged form possesses long electron spin relaxation and coherence times at room temperature[4]. Control of the NV-center spin state[5], as well as its initialization[6] and measurement[7], is implemented with external microwave and optical fields. In addition, NV-centers can be integrated into modern superconductor[8] or semiconductor[9] photonic systems, making them to be prominent candidates for large-scale solid-state quantum computations.

In past few years several prototypes of quantum information hybrid devices with NV-centers coupled to high-quality semiconductor optical resonators have been manufactured[10-14]. For example, NV-centers embedded in diamond nanocrystals interact with optical eigenmodes of microspheres[10], microdisks[11], microrings[12], waveguides[13] and photonic crystal defects[14]. The nanocrystals are disposed on the resonator surface by pick-and-place technique[15] and then fixed by adhesion. Experimental realizations of such interface show, however, very poor efficiency. In fact, the NV spectrum modification together with emission rate enhancement has been observed. The data obtained in Refs. 10-14 clearly point on achievement of weak coupling regime between NV-center(s) and cavity photon(s), while successful demonstration of strong coupling regime has not been reported yet. It may be explained by following reasons. First, the coupling coefficient $g$ of optical transition in a NV-center located at cavity surface to the field concentrated predominantly inside the cavity is rather weak. Second, diamond nanocrystals are proved to be imperfect containers for NV-centers because of their low optical properties. Third, the photon collection efficiency by confocal microscope remains quite low. Fourth, advanced schemes of Raman control adapted from atomic optics and applied to NV-centers do not work well. Fifth, the quality factors $Q = \omega_c/\kappa$ ($\omega_c$ is the frequency and $\kappa$ is the photon dissipation rate) of resonators fabricated by state-of-art technology are by several orders of magnitude lower than that required for strong coupling regime (i.e. when the condition $g \gg \gamma, \kappa$ is fulfilled). Note that the achievement of this regime is crucial for performance of quantum algorithms in the circuit quantum electrodynamics schemes.

*tsukanov@ftian.ru

Recent proposals[16-20] consider theoretical models aimed to remove or at least mitigate challenges listed above. All authors believe that NV-centers should be formed inside the body of diamond cavity[16] for coupling optimization rather than in surface-located nanocrystals. The key point is to synthesize NV-centers directly in the cavity field antinodes under high-level control of their position and number. The ordered array of NV-centers thus formed would be considered as the base for a large-scale solid-state quantum register. Unfortunately, recent attempts to reliably fabricate such a diamond photonic chip were unsuccessful because of technological imperfections inherent to ion-beam lithography[17-18]. Besides, current technique for individual NV-center synthesis at predetermined position in diamond matrix is still far from practical application. Therefore, given moderate quality of obtained devices, it would be very desirable to enhance quantum operation rates. Further, making use of photon chanelling in waveguide networks one could optimize the single-photon measurement. Another way makes use of sophisticated coupling schemes between NV-center, cavity and waveguide[19-20]. For example, switching between high-$Q$ cavity and low-$Q$ waveguide parts requires additional NV-center-doped cavity that functions as intermediate cell connecting the qubit with control and measurement parts of quantum computer.

Here, we theoretically investigate the possibility of speed-up of two-qubit transformations employing fast resonant optical scheme for interqubit coupling (fig.1). Such approach to the implementation of non-local entangling gates differs from those mentioned above. First, instead of simultaneous dispersive coupling of two (or more) qubits to common cavity mode, we tune qubits in exact resonance with the cavity sequentially (one by one). Namely, the control qubit has to be coupled with the cavity mode in the first turn. Given order in tunings enables us to make the evolution of target qubit to be conditioned by the state of control qubit. Obviously, if the control qubit being in definite logical state absorbs cavity photon from common waveguide, it provides controllable switching off of driving field for target qubit. Second, we apply fast resonant scheme for qubit-cavity interaction rather than slow off-resonant Raman scheme. This choice is provided by essentially two-level character of NV dynamics in single-mode approximation for cavity field without laser irradiation. Of course, we lose in this case the possibility to exploit three-level lambda scheme that provides noise suppression under off-resonant driving. With that, it would greatly simplify the control of the excitation process especially due to absence of polarized laser pulses. In addition, the number of resources as well as conditions imposed on system parameters is small for two-level system in comparison with more complex three-level case. Next, we do not need in laser irradiation as long as detuning control by electrostatic means is enough for our aims.

It is important to note that we use complex waveguide composed of sections. Each waveguide section is presented by linear chain of cavities (microdisks). It contains some number of NV-centers that we shall identify as a subregister. Neighbor waveguide sections are connected via switching NV-center that mediates photonic transfer between them being tuned in resonance with waveguide frequency. It allows us to consider isolated sections as moderate quality resonators with quality factor slightly lower than that of the single cavity. Therefore, coherent description of the photon in the section is still valid.

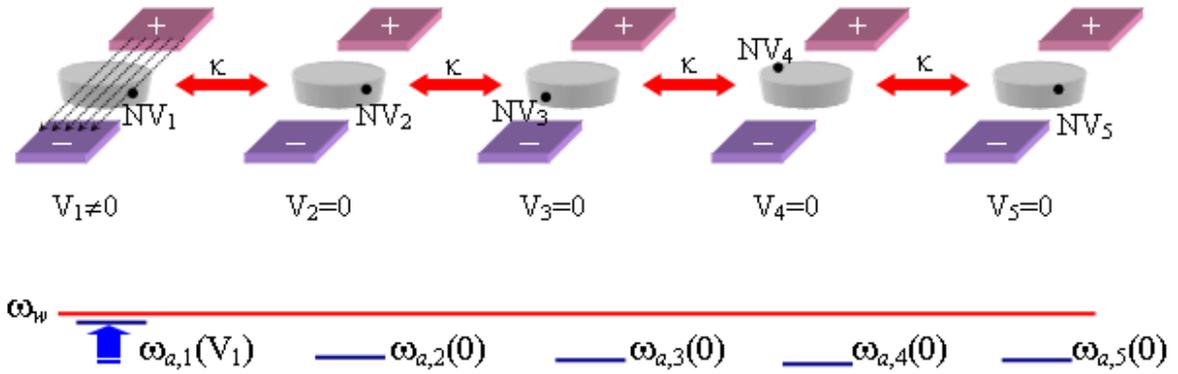

Figure 1. The scheme of the five-qubit quantum register based on quasi-linear photonic nanostructures consisting of disk diamond cavities coupled by the photon hopping $\kappa$. Each cavity contains the NV-center, encoding the quantum information in the electronic spin states. The optical transition frequency $\omega_{a,k}$ of the centers are controlled by a system of gates under individual electric bias $V_k$ ($k$ = 1 - 5). At the bottom is a diagram of the tuning of the first center in resonance with $\omega_w$ mode waveguide.

## 2. THE THEORETICAL SCHEME OF THE TWO-QUBIT OPERATION "CONTROLLED Z"

The idea of fast controlled-Z operation is as follows. Consider a two-level system which eigenstates are $|g\rangle$ (ground) and $|e\rangle$ (excited) and corresponding eigenenergies are $\varepsilon_g$ and $\varepsilon_e$. Let the system be placed in a high-quality single-mode resonator. As is well known, the two-level system coupled to a single quantized mode is described by the Jaynes-Cummings Hamiltonian ($\hbar = 1$)

$$H = \omega_w a^+ a + \omega_a |e\rangle\langle e| + g|g\rangle\langle e|a^+ + g^*|e\rangle\langle g|a, \tag{1}$$

where $\omega_w$ is the mode frequency, $\omega_a = \varepsilon_e - \varepsilon_g$ is the transition frequency of the system, $g$ is the coupling coefficient between mode and system, $a$ is the mode annihilation operator. As usual, the rotating wave approximation requiring that $|g| \ll \omega_w$ is applied in Eq. (1). We restrict ourselves by single-photon case and present the state vector in the form

$$|\Psi\rangle = c_0 \exp(-i\omega_w t)|g,1\rangle + c_1 \exp(-i\omega_a t)|e,0\rangle, \tag{2}$$

where $c_0$ and $c_1$ are the time-dependent probability amplitudes of combined electron-photon states $|g,1\rangle$ and $|e,0\rangle$. The evolution of the state vector depends strongly on the frequency detuning, $\delta = \omega_w - \omega_a$, and two opposite cases are possible. From one hand, by exact alignment of frequencies ($\delta = 0$) one obtains resonant evolution of $|\Psi(t)\rangle = U|\Psi(0)\rangle$ given by the matrix

$$U_{\delta=0}(\theta) = \begin{pmatrix} \cos\theta & -i\sin\theta \\ -i\sin\theta & \cos\theta \end{pmatrix}, \tag{3}$$

where $\theta = gt$ is the angle of Rabi rotation. From other hand, in the strongly-detuned case ($|\delta| \gg |g|$) the system undergoes to off-resonant dispersive evolution with matrix

$$U_{|\delta|\gg g}(\theta) = \begin{pmatrix} \exp(i\theta) & 0 \\ 0 & \exp(-i\theta) \end{pmatrix}, \tag{4}$$

where now $\theta = \dfrac{g^2 t}{\delta}$ is phase angle corresponding to Stark shifts of energy levels. For short time intervals ($t \ll |\delta|/g^2$) it is approximated by identity matrix with good accuracy. In what follows we shall be interested in three particular types of evolution described by matrices:

$$U_{\delta=0}(\pi/2) = \begin{pmatrix} 0 & -i \\ -i & 0 \end{pmatrix},\ U_{\delta=0}(\pi) = \begin{pmatrix} -1 & 0 \\ 0 & -1 \end{pmatrix},\ U_{|\delta|\gg g}(\theta) \approx \begin{pmatrix} 1 & 0 \\ 0 & 1 \end{pmatrix}.$$

The matrix $U_{\delta=0}(\pi/2)$ brings about the population inversion of system accompanied with the phase shift of $\pi/2$. The rotation $U_{\delta=0}(\pi)$ returns state populations to their initial values shifting the phase of the state vector by $\pi$. Obviously, $U_{\delta=0}(\pi) = U_{\delta=0}(\pi/2) U_{\delta=0}(\pi/2)$.

The Hamiltonian of two NV-centers that interact with common cavity mode is written

$$H = \omega_w a^+ a + \sum_{k=1,2} \left\{ \omega_{a,k}(t)|e_k\rangle\langle e_k| + D_{g,k}|+_k\rangle\langle +_k| + g_k|g_k\rangle\langle e_k|a^+ + g_k|g_k\rangle\langle e_k|a^+ \right\}. \tag{5}$$

In Eq. (5) the terms in brackets describe the three-level approximation for k-th NV-center where $|g_k\rangle \equiv |^3A, m=0_k\rangle$, $|e_k\rangle \equiv |^3E, m=0_k\rangle$, $|+_k\rangle \equiv |^3A, m=1_k\rangle$ are its relevant electron states, $\omega_{a,k}(t)$ is Stark-tunable frequency of optical transition $|g_k\rangle \leftrightarrow |e_k\rangle$ and $D_{g,k} = 2.87 \times 10^9$ Hz is the zero-field splitting. As is straightforwardly seen from Eq. (5), the

states $|g_k\rangle$ and $|e_k\rangle$ can vary their populations owing to the exchange by single quantum with cavity mode while the state $|+_k\rangle$ remains decoupled from the cavity. It means that the three-level approximation (5) is reduced further to the two-level approximation (1) for each center. However, we retain in Eq. (5) the terms $D_{g,k}|+_k\rangle\langle+_k|$ for clarity.

Let NV-centers be initially loaded in arbitrary superposition of their ground states whereas the cavity mode contains single photon. Four basis states, $|g_1,g_2,1\rangle$, $|g_1,+_2,1\rangle$, $|+_1,g_2,1\rangle$, and $|+_1,+_2,1\rangle$, span the two-qubit logical subspace. Remaining four states, $|e_1,g_2,0\rangle$, $|g_1,e_2,0\rangle$, $|e_1,+_2,0\rangle$, and $|+_1,e_2,0\rangle$, serve as auxiliary ones. The algorithm is divided into three steps. At first step, we organize resonant coupling of first NV-center (control qubit) with cavity mode to achieve $U_{\delta=0}(\pi/2)$ rotation. This operation requires the detuning $\delta_1(t)$ to be changed from maximum to zero, kept zero during the time interval $T_1(\pi/2) = \pi/2g_1$ and then returned to its maximum value. It is easy to verify that the logical states transform as $|g_1,g_2,1\rangle \to -i|e_1,g_2,0\rangle$, $|g_1,+_2,1\rangle \to -i|e_1,+_2,0\rangle$, $|+_1,g_2,1\rangle \to |+_1,g_2,1\rangle$, and $|+_1,+_2,1\rangle \to |+_1,+_2,1\rangle$. Namely, the control qubit is excited if and only if it has been loaded in the state $|g_1\rangle$. This step has two important consequences. From one hand, it results in phase accumulation of $\pi/2$ for two basis states. From other hand, it provides photon absorption by first NV-center leaving empty cavity before next step. At second step, second NV-center (target qubit) undergoes $U_{\delta=0}(\pi)$ rotation. Again, we choose the detuning pulse $\delta_2(t)$ to equalize frequencies $\omega_w$ and $\omega_{a,2}$ during the period $T_2(\pi) = \pi/g_2$. However, non-trivial evolution will take place only for the state $|+_1,g_2,1\rangle$ as long as it is the only state that couples to the active single-photon cavity at this step. As the result, one has $|+_1,g_2,1\rangle \to -|+_1,g_2,1\rangle$. Third step mimics the first one, and rotation $U_{\delta=0}(\pi/2)$ induces non-trivial evolution of basis states with first center excited. It retrieves photon to the cavity mode, returns control NV-center to the state $|g_1\rangle$ and supplies the logical states $|g_1,g_2,1\rangle$ and $|g_1,+_2,1\rangle$ with phase shift of $\pi/2$. Hence, their resulting phase shift is equal to $\pi$. As the result, the basis states $|g_1,g_2,1\rangle$, $|g_1,+_2,1\rangle$, and $|+_1,g_2,1\rangle$ accumulate phase shifts of $\pi$, whereas the basis state $|+_1,+_2,1\rangle$ remains unchanged. Thus, we realize the non-trivial two-qubit operation "–CZ". All steps are summarized below:

$$\begin{aligned}
|g_1,g_2,1\rangle &\xrightarrow{1,\pi/2} -i|e_1,g_2,0\rangle \xrightarrow{2,\pi} -i|e_1,g_2,0\rangle \xrightarrow{1,\pi/2} -|g_1,g_2,1\rangle \\
|g_1,+_2,1\rangle &\xrightarrow{1,\pi/2} -i|e_1,+_2,0\rangle \xrightarrow{2,\pi} -i|e_1,+_2,0\rangle \xrightarrow{1,\pi/2} -|g_1,+_2,1\rangle \\
|+_1,g_2,1\rangle &\xrightarrow{1,\pi/2} |+_1,g_2,1\rangle \xrightarrow{2,\pi} -|+_1,g_2,1\rangle \xrightarrow{1,\pi/2} -|+_1,g_2,1\rangle \\
|+_1,+_2,1\rangle &\xrightarrow{1,\pi/2} |+_1,+_2,1\rangle \xrightarrow{2,\pi} |+_1,+_2,1\rangle \xrightarrow{1,\pi/2} |+_1,+_2,1\rangle
\end{aligned} \quad (6)$$

We simulate this gate by solving numerically the time-dependent Schrodinger equation with Hamiltonian (5). The simulation results a presented in figs 2 – 4. The optical transition frequencies of NV-centers are defined by rectangle Stark bias pulses (fig. 2). According to Ref. 19 we use the values $\omega_{a,k}(0) = 2.95 \times 10^{15}$ Hz for the unperturbed zero-phonon line frequency, $g_1 = 10^{10}$ Hz and $g_2 = 0.9 \times 10^{10}$ Hz for the coupling coefficients, and $\delta_k(0) = \omega_w - \omega_{a,k}(0) \sim 10^{12}$ Hz for the maximum detuning. All values in figs are in $\omega_{a,k}(0)$ units. Populations of logical states $p_{00}$, $p_{01}$, $p_{10}$ and $p_{11}$ evolve in accordance with the predictions of a simple two-level model (fig. 3). As seen from fig. 4, the first phase of the three states at the end of the Stark pulse changes to $\pm\pi$, while the fourth phase of the state does not change. Thus, the numerical analysis confirms our arguments and proves the validity of the developed scheme.

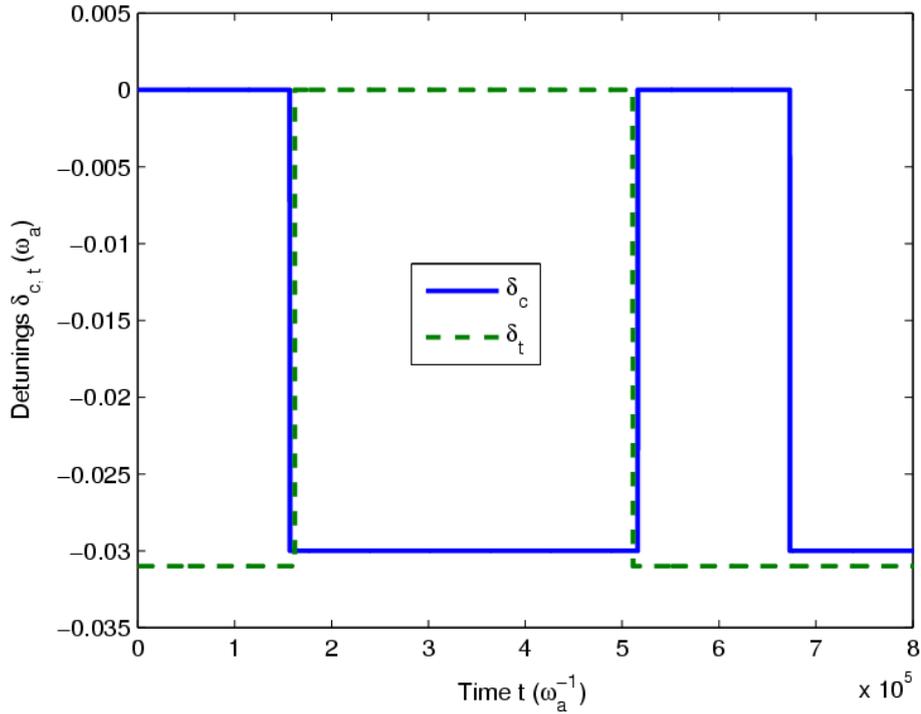

Figure 2. The frequency detunings $\delta_c(t)$ and $\delta_t(t)$ in units of $\omega_{a,k}(0)$ for the control (c) and the target (t) qubits vs time in units of $\omega^{-1}_{a,k}(0)$.

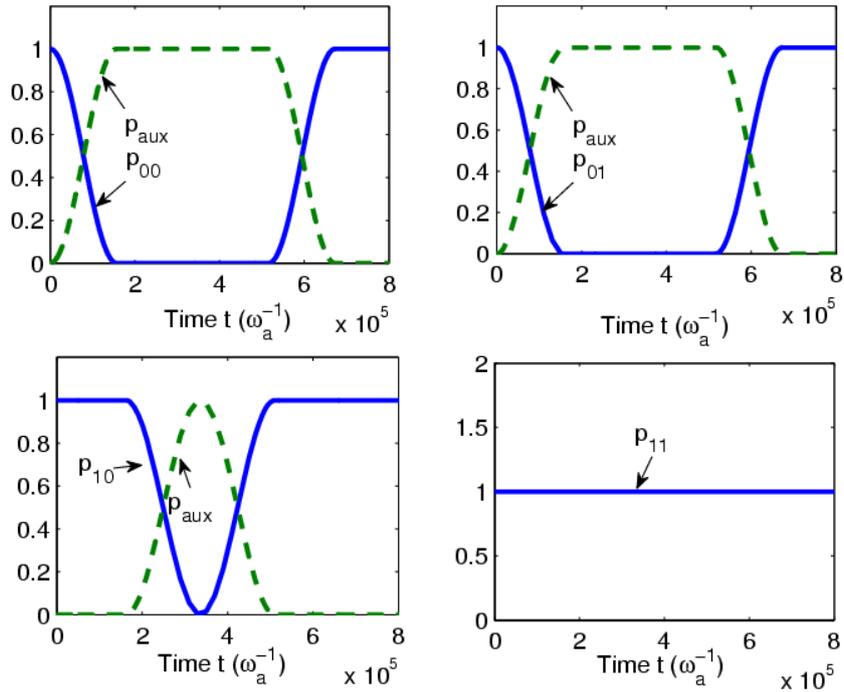

Figure 3. The populations $p_{00}$, $p_{01}$, $p_{10}$, $p_{11}$ of logical states and the populations $p_{aux}$ of auxiliary states of NV+cavity system vs time in units of $\omega^{-1}_{a,k}(0)$.

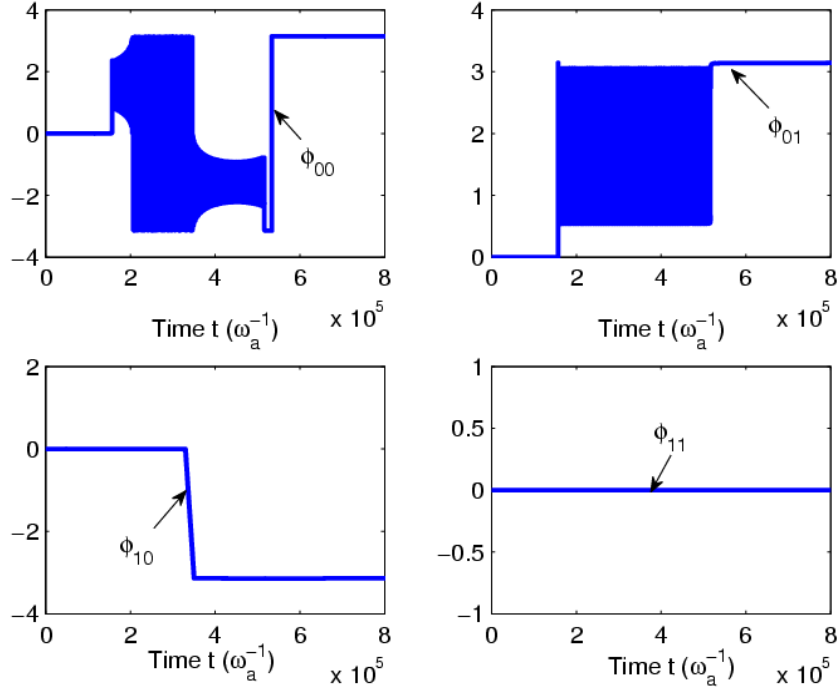

Figure 4. The phases of logical states vs time in units of $\omega^{-1}_{a,k}(0)$.

## 3. THE MODEL OF A WAVEGUIDE RESONATOR AND SPECTRAL CHARACTERISTICS FOR A ONE-DIMENSIONAL MICRODISK CHAIN

In order to organize an interaction and quantum correlations between the states of two NV-centers one can use a delocalized photon mode of a cavity chain. In our model the microdisk cavities act as containers for NV-centers. Let us consider a general model to find the eigenmodes and the electromagnetic field distribution of a high-$Q$ cavity array[21]. The array consists of the one-dimensional chain of the identical microdisk cavities with radius $R$ and thickness $h$ (fig. 1). Suppose that the interaction between their individual optical modes formed by a photon hopping is weak, so one can use the approach similar to tight-binding approximation in solid state physics. We represent the electric field of the microdisk chain eigenmode as a linear combination of single disks modes:

$$\mathbf{E}_\Omega(\mathbf{r}) = \sum_p \exp(iKpL)\mathbf{E}_\omega(\mathbf{r} - pL\mathbf{e}_x), \qquad p = 0, \pm 1, \pm 2 \ldots \qquad (7)$$

where $K$ is a photon quasiwave vector in the disk chain, $p$ is a disk number, $L$ is a distance between the disk centers, $\mathbf{e}_x$ is the unit vector along the chain direction. Since $\mathbf{E}_\Omega$ satisfies the Bloch theorem, one can restrict the $K$ value with the first Brillouin zone, i.e. $-\pi/L \leq K \leq \pi/L$. $\mathbf{E}_\Omega(\mathbf{r})$ и $\mathbf{E}_\omega(\mathbf{r})$ satisfy the Maxwell equation

$$\mathbf{rot\,rot\,E}_\Omega = \frac{\Omega^2}{c^2} n^2(\mathbf{r})\mathbf{E}_\Omega(\mathbf{r}), \qquad \mathbf{rot\,rot\,E}_\omega = \frac{\omega^2}{c^2} n_0^2(\mathbf{r})\mathbf{E}_\omega(\mathbf{r}), \qquad (8)$$

where $n(\mathbf{r})$ and $\Omega$ are a refractive index and a mode frequency of the cavity chain, $\omega$ and $n_0(\mathbf{r})$ are a refractive index and a mode frequency of the single microdisk, $n_0(\mathbf{r}) \equiv n_c$ inside the disk, $n(\mathbf{r}) = n_0(\mathbf{r}) \equiv 1$ outside the cavity. Substituting (7) into (8), one obtains

$$\Omega^2 = \omega^2 \frac{\sum_p \exp(iKpL)\beta_p}{\sum_p \exp(iKpL)\alpha_p} = \omega^2 \frac{\beta_0 + \sum_{p\neq 0} \exp(iKpL)\beta_p}{\beta_0 + \Delta\alpha + \sum_{p\neq 0} \exp(iKpL)\alpha_p}, \qquad (9)$$

where $\alpha_p$, $\beta_p$ и $\Delta\alpha$ are defined as

$$\begin{aligned}
\alpha_p &= \int d\mathbf{r}\, \mathbf{E}_\omega(\mathbf{r}) n^2(\mathbf{r}) \mathbf{E}_\omega(\mathbf{r} - pL\mathbf{e}_x), & p \neq 0 \\
\beta_0 &= \int d\mathbf{r}\, \mathbf{E}_\omega^2(\mathbf{r}) n^2(\mathbf{r}) \\
\beta_p &= \int d\mathbf{r}\, \mathbf{E}_\omega(\mathbf{r}) n_0^2(\mathbf{r} - pL\mathbf{e}_x) \mathbf{E}_\omega(\mathbf{r} - pL\mathbf{e}_x), & p \neq 0 \\
\Delta\alpha &= \int d\mathbf{r}\, \mathbf{E}_\omega^2(\mathbf{r}) \left[ n^2(\mathbf{r}) - n_0^2(\mathbf{r}) \right].
\end{aligned} \qquad (10)$$

If the coupling between the cavities is sufficiently weak, we can take into account only the interaction between the nearest neighbors, i.e. $\alpha_p = 0$ and $\beta_p = 0$ if $p \neq \pm 1$. The symmetry of our system means that $\alpha_1 = \alpha_{-1}$, $\beta_1 = \beta_{-1}$. Finally we assume $\alpha_1 \ll \beta_0$, $\beta_1 \ll \beta_0$, $\Delta\alpha \ll \beta_0$. Then expression (9) is reduced to a simple dispersion relation:

$$\Omega = \omega \left( 1 - \frac{\Delta\alpha}{2\beta_0} - \frac{\zeta}{\beta_0} \cos KL \right), \qquad (11)$$

where

$$\zeta = \alpha_1 - \beta_1 = \int d\mathbf{r}\, \mathbf{E}_\omega(\mathbf{r}) \left[ n^2(\mathbf{r}) - n_0^2(\mathbf{r} - L\mathbf{e}_x) \right] \mathbf{E}_\omega(\mathbf{r} - L\mathbf{e}_x)$$

The coupling between neighbor cavities is equal to

$$\kappa = \zeta \frac{\omega}{\beta_0} \qquad (12)$$

Therefore, each mode of the single microdisk with frequency $\omega$ splits into a band of $2\kappa$ width, delocalized throughout the chain. Note that the electrical field integrating contained in (11) is contributed by area inside the disks only and this circumstance simplifies the calculations significantly.

The first step to find the electric field and the spectrum of one-dimensional cavity chain is the calculation of the eigenmode frequency and the electrical field distribution of the single microdisk because these values are included in the expressions (9) – (12). In this paper we studied the qubits based on the diamond NV-centers, where the wavelength of the transitions between electron spin states $|g\rangle$ and $|e\rangle$ corresponds to the frequency of zero-phonon line $\lambda_0 = 637$ nm. Therefore, in order to implement the quantum operations one should choose the disk size (thickness $h$ and radius $R$) so that the wave length of one of its mode is close to $\lambda_0$. The distribution of the axial component $E_z$ of the electric field for TM modes meets the wave equation, written in cylindrical coordinates

$$\frac{\partial^2 E_z}{\partial \rho^2} + \frac{1}{\rho}\frac{\partial E_z}{\partial \rho} + \frac{1}{\rho^2}\frac{\partial^2 E_z}{\partial \varphi^2} + \frac{\partial^2 E_z}{\partial z^2} + k^2 n_c^2 E_z = 0 \qquad (13)$$

where $k = \omega/c$ is a wave vector, $\omega$ is the eigenfrequency, $c$ is a speed of light in vacuum, $n_c = 2.4$ is a diamond refractive index. Introducing an effective refractive index $\tilde{n}$ and using the boundary conditions for the electromagnetic field at the disk boundaries[22,23] the analytical solution of (13) is given by a system of equations:

$$\tilde{n} \frac{J_{m+1}(k\tilde{n}R)}{J_m(k\tilde{n}R)} = \frac{H_{m+1}^{(1)}(kR)}{H_m^{(1)}(kR)}, \qquad (14)$$

$$\sqrt{n_c^2 - \tilde{n}^2}\, \text{tg}\, \frac{\beta h}{2} = n_c^2 \sqrt{\tilde{n}^2 - 1}, \qquad \beta = k\sqrt{n_c^2 - \tilde{n}^2} \qquad (15)$$

where $J_m$ is a Bessel function of first kind, $H_m^{(1)}$ is a Hankel function of first kind, $m = 0, \pm1, \pm2, \ldots$ is an azimuthal number. Since the qubits are located near the edge of microdisk, the mode antinodes are to locate along the edge of the disk. TM microdisk modes with the large $m$ (the so-called "whispering gallery modes") satisfy this requirement. Solving the equation system (14), (15), one can find the structural parameters $R$ and $h$ for the fixed eigenvector $k$, corresponding to the photon wavelength $\lambda_0$. The axial electric field distribution of the TM mode is given by the following expression

$$E_\omega(\mathbf{r}) \equiv E_z(\mathbf{r}) = F(\rho)\exp(i\beta z + im\varphi), \qquad F(\rho) = \begin{cases} \dfrac{J_m(k\tilde{n}\rho)}{J_m(k\tilde{n}R)}, & \rho < R \\ \dfrac{H_m^{(1)}(k\rho)}{H_m^{(1)}(kR)}, & \rho > R. \end{cases} \qquad (16)$$

Table 1 shows the microdisk size parameters for which the wavelength of the whispering gallery mode is $\lambda_0 = 637$ nm, and figs 5 (a) and 5 (b) represents of the electric field amplitude $F$ of the TM$_{40,\,1}$ ($m = 40$) and TM$_{50,\,1}$ ($m = 50$) modes versus the distance $\rho$ from the disk center for the cavities of different radius $R$ and thickness $h$. For the calculated microdisk size (Table 1) NV-centers located in the mode antinode are coupled efficiently. Since the mode wavelength is fixed and equal to $\lambda_0$ the disk with larger $R$ should have a smaller thickness $h$. In this case for cavities with sufficiently large radius, characterized by low values of radiative quality, the field is large sufficiently and decays slowly far away from the microdisk as $F(\rho) \propto H_m^{(1)}(k\rho) \propto \exp(ik\rho)/\sqrt{k\rho}$. Thus, the above approach to the calculation of the microdisk chain spectrum becomes incorrect. However, if the disk is located near the edge of the neighbor one in so-called evanescent region where the field decays exponentially, one can believe that the weak coupling between two such cavities is valid.

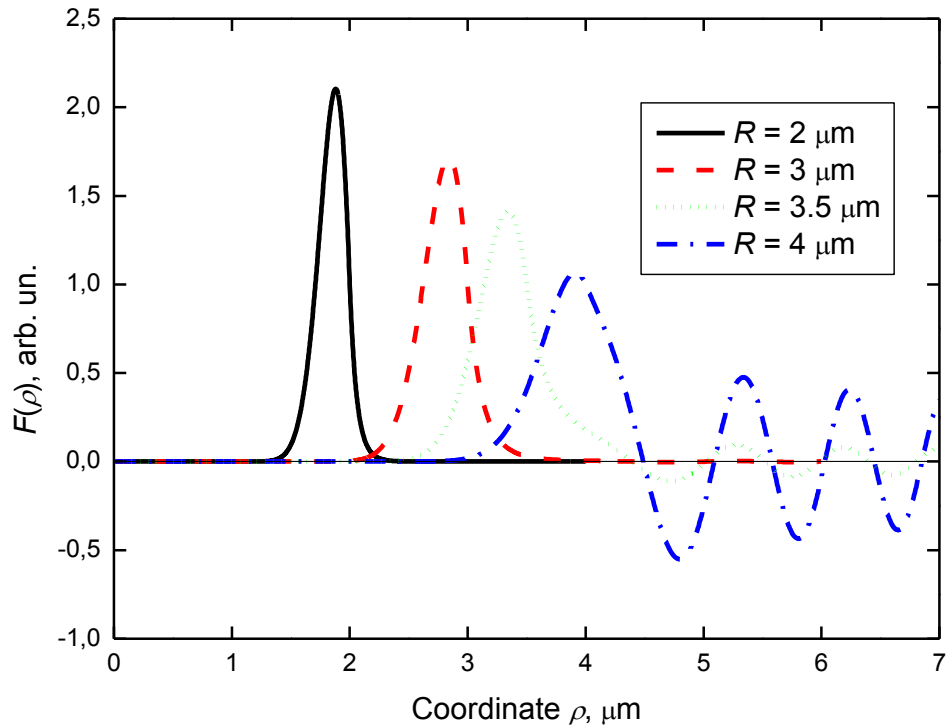

Figure 5 (a). The electric field amplitude $F$ of the TM$_{40,\,1}$ mode vs the distance $\rho$ from the disk center. Solid line - $R = 2$ μm, dashed line - $R = 3$ μm, dotted line - $R = 3.5$ μm, dash-dotted line - $R = 4$ μm.

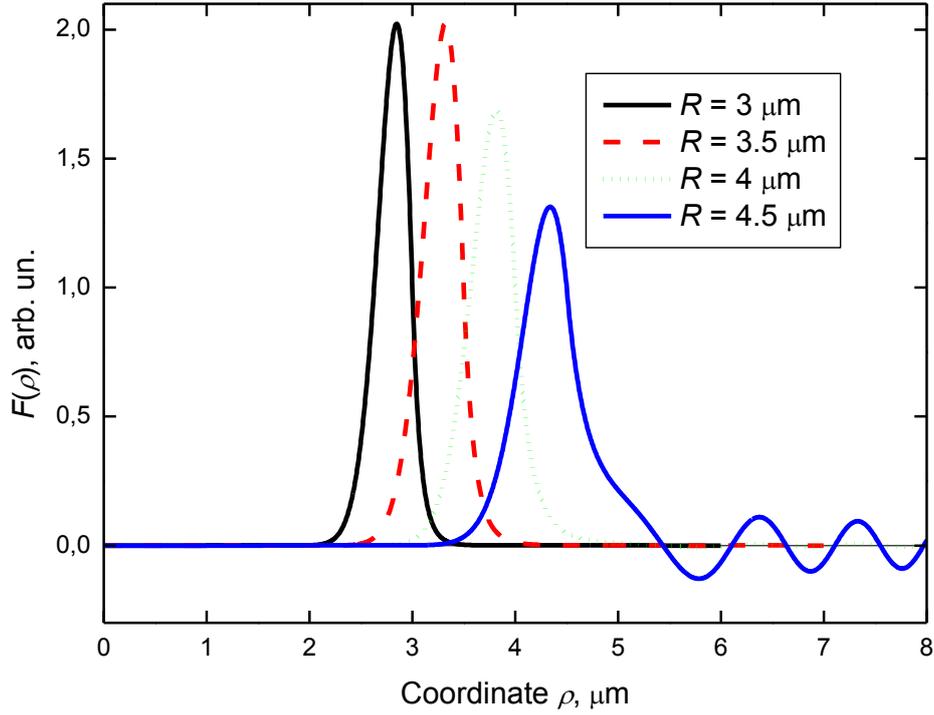

Figure 5 (b). The electric field amplitude $F$ of the $TM_{50,1}$ mode vs the distance $\rho$ from the disk center. Solid line - $R = 3$ μm, dashed line - $R = 3.5$ μm, dotted line - $R = 4$ μm, dash-dotted line - $R = 4.5$ μm.

Table 1. Size of the diamond ($n = 2.4$) microdisk cavity for which the photon wavelength of TM mode is equal to $\lambda_0 = 637$ nm

| $m = 40$ | |
|---|---|
| $R$, μm | $h$, μm |
| 2.0 | 0.469 |
| 2.5 | 0.185 |
| 3.0 | 0.143 |
| 3.3 | 0.128 |
| 3.5 | 0.118 |
| 3.7 | 0.108 |
| 4.0 | 0.088 |

| $m = 50$ | |
|---|---|
| $R$, μm | $h$, μm |
| 2.5 | 0.397 |
| 3.0 | 0.194 |
| 3.5 | 0.153 |
| 3.7 | 0.143 |
| 4.0 | 0.130 |
| 4.5 | 0.111 |
| 5.0 | 0.085 |

Having calculated the spectrum and the electric field of the single cavity, we pass to the influence of the size and distance between the equal microdisks on the coupling $\kappa$ (12) between neighbouring cavities. Fig. 6 shows the landscape electric field profile of the microdisk chain with radius $R = 3$ μm for $K L = 17 \pi/40$ and the distance between the disk centers $L = 2.2 R$. $TM_{40,1}$ mode antinodes, one of which has the NV-center, are located along the edge of each cavity (the number of antinodes corresponds to the azimuthal number $m$). The field amplitude is small outside microdisk, and we assume that for such configuration of the one-dimensional waveguide the neighbouring microdisk coupling coefficient $\kappa$ is exponentially decreases with increasing of the distance between disk centers. Fig. 7(a) shows the dependence of the logarithm of the ratio of the coupling $\kappa$ to the energy of the NV-center zero-phonon line $E_0 = 1.945$ eV for $TM_{40,1}$ mode on the distance $L$ between the microdisks at different values of the cavity radius $R$. The ratio is seen almost linear and, therefore, the coupling decreases exponentially with increasing $L$. The $\kappa$ reduce becomes slower when the disk radius increases because of the field enhancement outside the cavities. Similar results are obtained for the $TM_{50,1}$ mode (fig. 7(b)). Thus, one can suggest that the tight-binding approximation[21] described above is valid for the considered range of $R$ and $L$. Comparison of the results for $m = 40$ and $m = 50$ shows that for TM modes with larger $m$ coupling $\kappa$ between modes in the cavity chain is weaker, since the amplitude of the field outside microdisks is smaller as well. Implementation of quantum operations requires the inequality $\kappa \gg g_1, g_2$ to be satisfied, therefore, the coupling between the cavities should be in the range $10^{-4}$ - $10^{-5}$ eV. This choice of structure parameters provides effective photon hopping between the neighbouring microdisks and keeps the validity of the interaction cavity approach used in this paper. We calculated the dependence of $\kappa$ on the distance $L$ between the centers of neighboring disks (see Tables 2 and 3) at different values of $R$ for $TM_{40,1}$ and $TM_{50,1}$ modes. It is seen that for $TM_{40,1}$ mode one should choose the distance in a range $L/R$ = 2.11 - 2.21, however for $R$ = 3 μm the ratio $L/R$ = 2.31, i.e. disks with larger radius may be placed further away from each other. In addition, for $TM_{50,1}$ mode $L$ should be significantly smaller than for $TM_{40,1}$ mode.

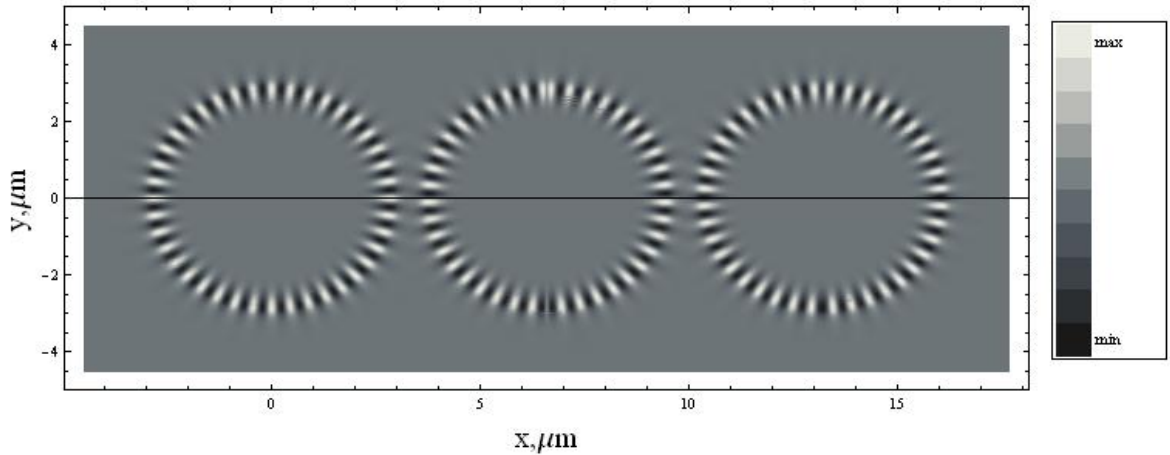

Figure 6. The electric field profile of the mickrodisk chain of radius $R$ = 3 μm for the distance between the disks $L = 2.2 R$, $K L = 17 \pi/40$.

Table 2. The dependence of coupling $\kappa$ between neighboring cavities on the distance $L$ at different values of $R$ for $TM_{40,1}$ mode

| $L/R$ | $R = 2$ μm | $R = 2.5$ μm | $R = 3$ μm |
|---|---|---|---|
| 2.01 | 5.4162318 | 6.1629656 | 7.3221493 |
| 2.11 | 0.19195095 | 0.31170879 | 0.60936661 |
| 2.21 | $8.4024303 \cdot 10^{-3}$ | $2.0020717 \cdot 10^{-2}$ | $6.8041825 \cdot 10^{-2}$ |
| 2.31 | $4.4752246 \cdot 10^{-4}$ | $1.6128662 \cdot 10^{-3}$ | $1.0331791 \cdot 10^{-2}$ |
| 2.41 | $2.8629954 \cdot 10^{-5}$ | $1.6140768 \cdot 10^{-4}$ | $2.0305286 \cdot 10^{-3}$ |
| 2.49 | $3.5948418 \cdot 10^{-6}$ | $2.9785056 \cdot 10^{-5}$ | $8.0271189 \cdot 10^{-4}$ |

Table 3. The dependence of coupling $\kappa$ between neighboring cavities on the distance $L$ at different values of $R$ for $TM_{50,1}$ mode

| $L/R$ | $R = 2.5$ μm | $R = 3$ μm | $R = 3.5$ μm |
|---|---|---|---|
| 2.01 | 3.7089467 | 4.1999312 | 4.7383159 |
| 2.11 | $5.7043459 \cdot 10^{-2}$ | $9.0470660 \cdot 10^{-2}$ | 0.15825279 |
| 2.21 | $1.1423516 \cdot 10^{-3}$ | $2.6025582 \cdot 10^{-3}$ | $7.3626257 \cdot 10^{-3}$ |
| 2.31 | $2.9232764 \cdot 10^{-5}$ | $9.8328396 \cdot 10^{-5}$ | $4.7248165 \cdot 10^{-4}$ |
| 2.41 | $9.4061343 \cdot 10^{-7}$ | $4.8134632 \cdot 10^{-6}$ | $4.1644441 \cdot 10^{-5}$ |
| 2.49 | $7.0305704 \cdot 10^{-8}$ | $5.1537858 \cdot 10^{-7}$ | $7.5212790 \cdot 10^{-6}$ |

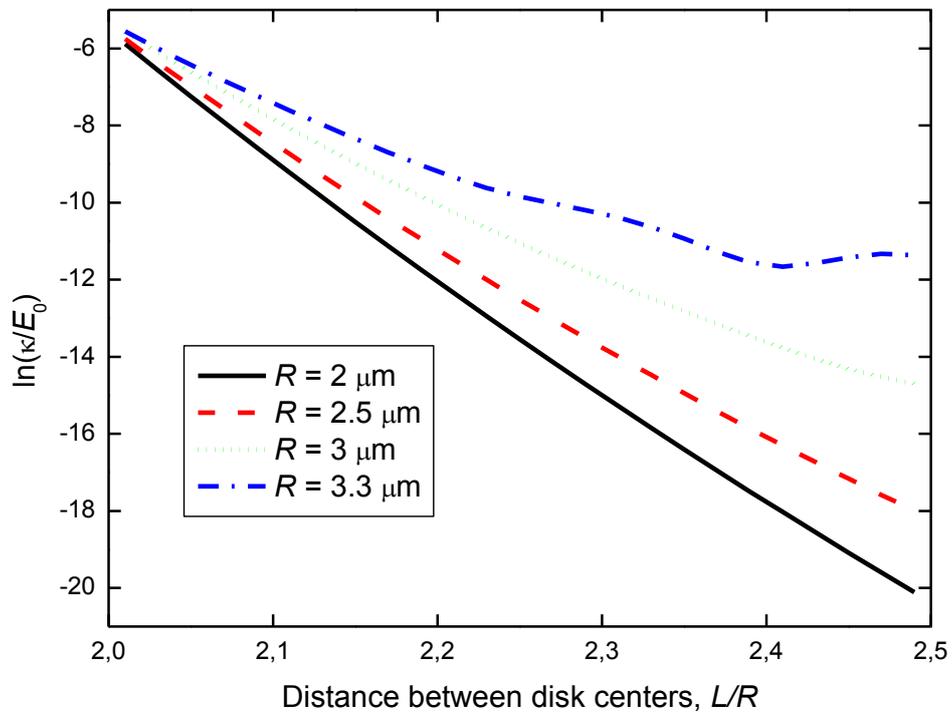

Figure 7 (a). The dependence of the logarithm of the ratio of the coupling $\kappa$ to the energy of the NV-center zero-phonon line for $TM_{40,1}$ mode on the distance $L$ between the microdisks. Solid line - $R = 2$ μm, dashed line - $R = 2.5$ μm, dotted line - $R = 3$ μm, dash-dotted line - $R = 3.3$ μm.

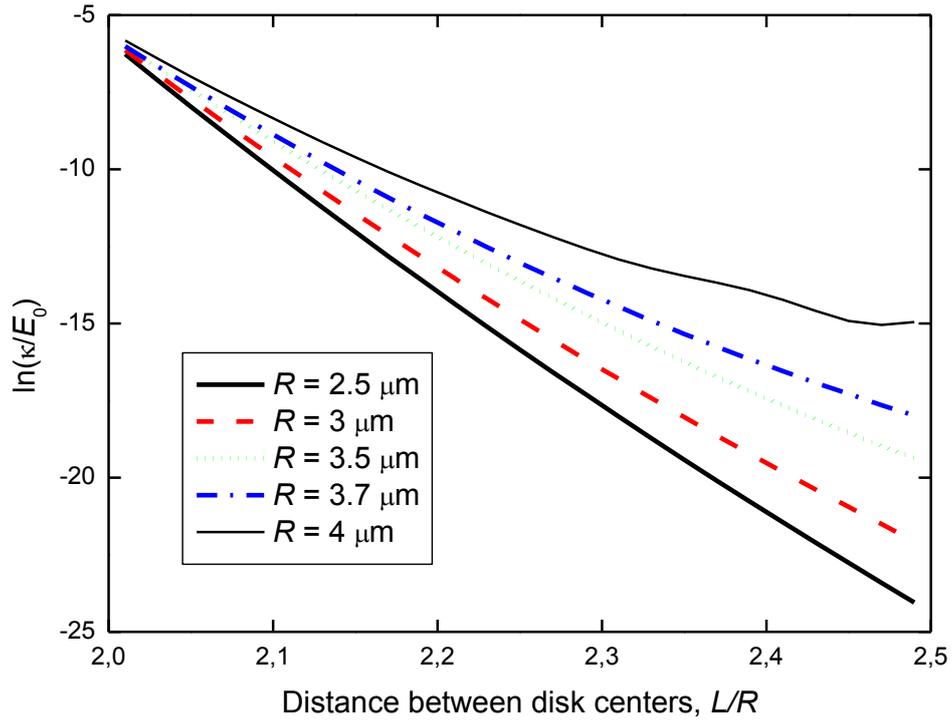

Figure 7 (b). The dependence of the logarithm of the ratio of the coupling $\kappa$ to the energy of the NV-center zero-phonon line for $TM_{50,1}$ mode on the distance $L$ between the microdisks. Solid line - $R = 2.5$ μm, dashed line - $R = 3$ μm, dotted line - $R = 3.5$ μm, dash-dotted line - $R = 3.7$ μm, thin line - $R = 4$ μm.

## 4. CONCLUSIONS

In this paper we propose the scheme of the coherent interaction of NV-centers in diamond nanostructures represented by the one-dimensional chain of optical microdisk resonators. Theoretical analysis of the NV-center electronic states indicates on the possibility of the realization of fast two-qubit operations (e.g. CZ) by external control of their optical transition frequencies. As we show it gives rise to fine tuning of the selected centers into resonance with a common photon mode in the cavity-based structure. The tuning is performed by applying of bias voltage (Stark effect) to individual control gates attached to each NV-center.

Using computer simulations we calculated the frequency and the electric field distribution of the common photon mode (the combination of the WGM modes of the diamond microdisk cavities arranged in a one-dimensional chain). The size of a single cavity provided the effective optical coupling in the system "NV-center + microdisk" is obtained.